# BÖLÜM 5

## MODELLEME VE SİMÜLASYON

Dr. Serdar ABUT

[1] Siirt Üniversitesi, Bilgisayar Mühendisliği Bölümü, Siirt, serdarabut@gmail.com

# 1. GİRİŞ

Bilgisayar modellemesi ve simülasyonu, sistem davranışlarının analiz etmek ve tanımlayıcı veya tahmine dayalı modlarda işleyişindeki stratejileri değerlendirmek için kullanılmaktadır (Abar ve ark., 2017). Model kavramı, halihazırda var olan veya henüz planlanmış belirli bir gerçekliğin soyut ve basitleştirilmiş bir temsili olarak kabul edilmektedir. Modeller, gözlemlenen bir olguyu incelemek ve açıklamak veya gelecekteki olguyu öngörmek için yaygın olarak kullanılmaktadır. Bilgisayar simülasyonu terimi, modelleme seçimlerinin sonuçlarını öngörerek karmaşık bir sistemin davranışına (örneğin biyolojik veya sosyal sistemler) dönük tahmin elde etmek için bir hesaplama modelinin kullanılması, aynı zamanda tasarımları ve planları gerçek dünyada(örn. mimari tasarımlar, yollar veya trafik ışıkları) fiilen hayata geçirmeden değerlendirmek anlamına gelmektedir. Simüle edilen sistemin fiilen gözlemlenemediği, etik (örneğin, insanların güvenliği söz konusu olacaksa) veya deney ve veri toplamanın maliyetinin fazla olması gibi nedenlerle çoğu zaman bu "sentetik ortamların" kullanımı gerekmektedir (Bandini ve ark., 2009).

1960'larda ilk olarak geliştirilen modelleme ve simülasyon (M&S) teorisi, başlangıçta dinamik sistemleri temsil edecek matematiksel gösterimler sağlamayı amaçlamıştır. Sistem teorik olarak, sistem yapısı (bir sistemin iç yapısı) ve davranışı (dış tezahürü) arasında ayrım yapar. Bir kara kutu olarak bakıldığında (Şekil 1), bir sistemin dışa dönük davranışı, girdi katmanına aldığı değerler ile çıktı katmanında üretilen değerler arasında kurulan ilişkidir. Sistem yapısını bilmek, davranışını anlamamızı (analiz etmemizi, simüle etmemizi) sağlar. Modelleme ve simülasyon işleminin temel amacı gözlemlenen bir davranışın geçerli bir temsilini keşfetmektir.

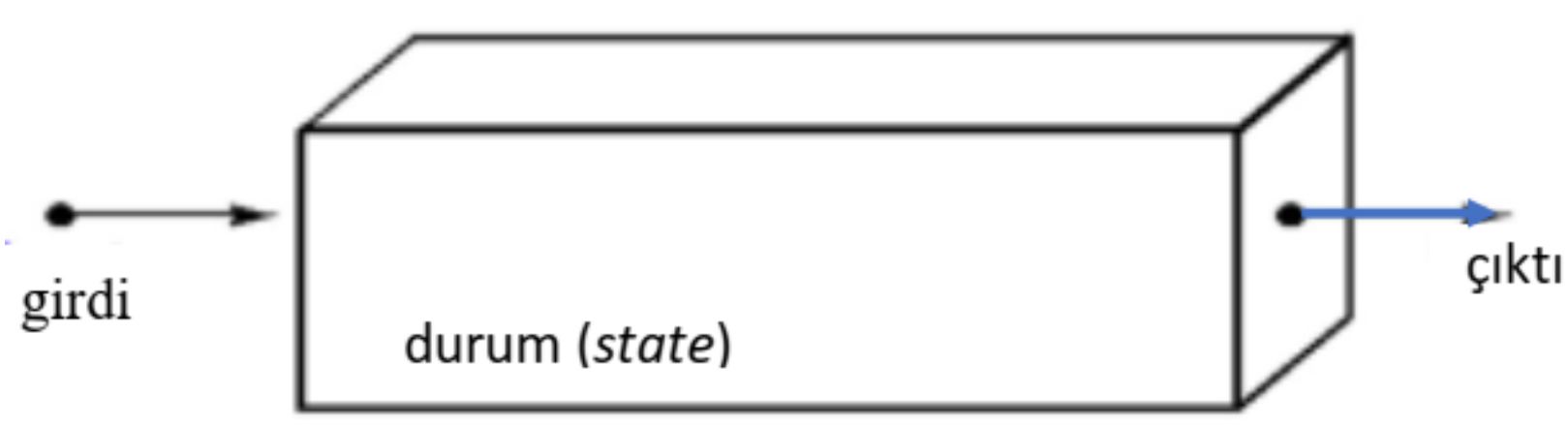

**Şekil 1:** Temel bir sistemin bileşenleri

Zaman ve durum parametrelerinin kesikli ve sürekli olmasına göre denklem sistemleri farklılık gstermektedir. Geleneksel diferansiyel denklem sistemlerinde (*Differantial Equation System* -DES) zaman ve durumlar(*states*) süreklidir. Fakat, otomatlar gibi kesikli bir zaman temelinde çalışan sistemler kesikli zamanlı sistemler (*Discrete Time System* - DTS) olarak isimlendirilmektedir. Bu iki sistem, Newton-Leibniz döneminden beri matematiksel gösterimler sunmaya devam etmektedir. Üçüncü bir yaklaşım olarak, büyük ölçüde algoritmalar ve simulasyon dillerinin etkisi altında kalan ayrık olay modelleri (*Discrete Event Systems* - DEVS) ileri sürülmüştür. Kontrol ve tasarımda soyutlamanın faydaları görüldükçe bu formdaki mutant modellerin tasarımı yaygınlaşmıştır (Ho, 1992).

Kompleks bir sistemin gerçekçi bir modelleme ve simülasyonu, sistem ve çevrenin deterministik olmayan (*nondeterministic*) özelliklerini muhakkak içermelidir. 'Deterministik olmayan' teriminden kasıt, sistemin tepkilerinin kesin bir şekilde tahmin edilemezliğidir. Bu belirsizlik, sistem ve çevrenin ürettiği belirsizlik veya sistem ile insan etkileşiminden doğan belirsizliklerdir. Nondeterminizm kavramı, sistemin anormal ya da sıradışı ortamlardaki çalışmasının güvenliğini tayin etmek için risk yönetimi alanında büyük bir yer edinmiştir. Nükleer raktör analizi, çevresel etki analizi ve deprem mühendisliği gibi alanlarda nondeterministik etkiler ölçülmeye çalışılmaktadır. Risk yönetimi alanındaki temel kurgu, matematiksel modeller aracılığıyla parametre belirsizliklerini modellemeye ve yayılımını incelemeye yöneliktir (Oberkampf ve ark., 2002).

Reaktör güvenliğini sağlamak amacıyla, anormal veya arıza senaryolarında, istenmeyen olayın meydana gelebileceği tüm güvenilir yolları bulmak için tasarlanan hata ağaçlarının (Vesely ve ark., 1987) analiz edilmesinde belirsizliklerin etkisinin tayin edilebilmesine büyük önem verilmiştir. Oberkampf ve ark. çalışmalarında, belirsizlik kaynaklarını tayin edebilmek için olasılık yöntemlerini kullanmışlardır. Modelleme ve simülasyonda hata ve belirsizliği belirlemeye yönelik gerçekleştirdikleri çalışmalarında, tahmin edilen belirsizlik kaynakları, deterministik bir model aracılığıyla Monte Carlo örneklemesi gibi örnekleme yöntemlerini kullanarak simüle edilmiştir (Oberkampf ve ark., 2002).

Sistem mühendisliği ve yöneylem araştırması toplulukları, modelleme ve simülasyon için birçok genel ilke ve prosedür geliştirmiştir. Bu konuda çalışan araştırmacılar, modelleme ve simülasyonun çeşitli aşamalarını tanımlama ve kategorize etme konusunda önemli ilerleme kaydetmişlerdir (Bossel, 2018; Zeigler, 2018; Neelamkavil, 1987; Law, 2000; Banks, 1998). Yöneylem araştırmalarında temel odak noktası, sorunlu varlığın tanımlanması, kavramsal modelin tanımlanması, veri ve bilgi kalitesinin değerlendirilmesi, doğrulama metodolojisi ve karar vermede yardımcı olarak simülasyon sonuçlarının kullanılmasıdır (Oberkampf ve ark., 2002).

1979'da, Bilgisayar Simülasyonu Topluluğu Model Güvenilirliği Teknik Komitesi, modelleme ve simülasyonun birincil aşamalarını ve faaliyetlerini tanımlayan bir diyagram geliştirmiştir (Şekil 2). Şekil 2'de görüldüğü gibi analiz aşaması, gerçeğin kavramsal bir modelini oluşturmak için kullanılmaktadır. Programlama aşaması ile kavramsal model, hesaplanan modele dönüştürülür. Daha sonra gerçekliği simule etmek için bilgisayar simulasyonu kullanılır. Düz ve basit olmasına rağmen Şekil 2' deki diagram, modelleme ve simulasyonun iki temel aşamasının birbiri ile ve gerçeklikle ilişkisini göstermektedir. Şekil 2'de ayrıca model yeterliliği, model geçerliği ve model doğrulama aşamaları da gösterilmektedir. Fakat Şekil 2'de, sistemi tanımlayabilecek parçalı diferansiyel eşitliklerin (*Partial Differential Equations*-PDEs) çözümü için gereken ayrıntılı faaliyetleri veya belirsizliğin tahmini için gerekli faaliyetleri ele almamaktadır.

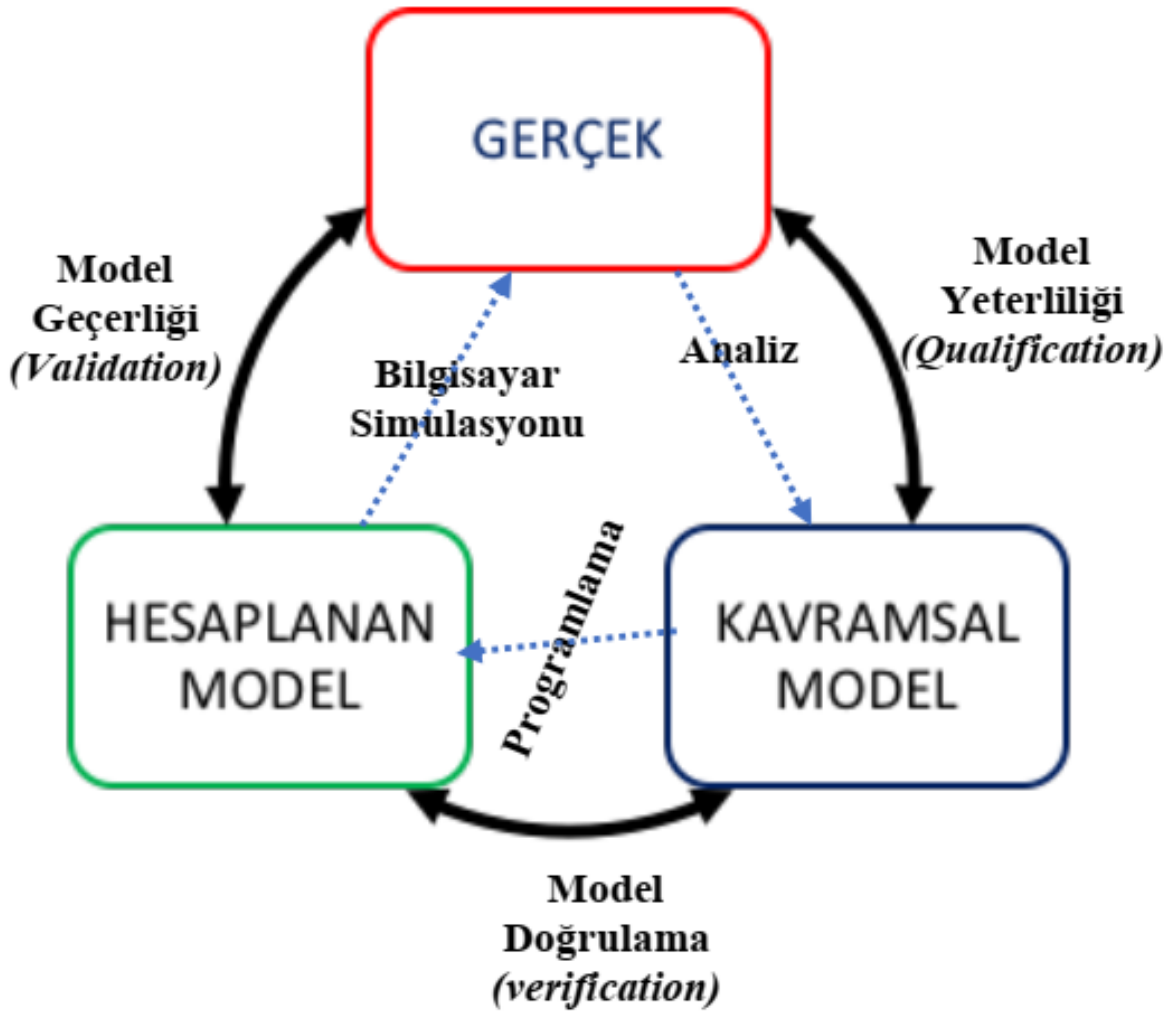

**Şekil 2:** Bilgisayar Simulasyon Topluluğu tarafından tasarlanan modelleme ve simulasyon şeması (Schlesinger, 1979).

Jacoby ve Kowalik, 1980'de modelleme ve simülasyon aşamaları için daha ayrıntılı bir tasarım önermişlerdir (Şekil 3) (Jacoby ve ark., 1980). Çalışmalarında, sadece modelleme ve simülasyon aşamalarını daha iyi tanımlamakla kalmayıp, aynı zamanda sürecin matematiksel modelleme yönlerini de öne çıkarmışlardır. Modelleme çalışmasının amacı ve hedefi netleştirildikten sonra bir prototip modelleme çalışması yürütülmektedir. Ön modelleme ve matematiksel modelleme aşamalarında çeşitli alternatif matematiksel modeller oluşturulmakta ve uygulanabilirlikleri değerlendirilmektedir. Çözüm tekniği aşamasında, matematiksel modeli veya modelleri çözmek için sayısal yöntemler belirlenmektedir. Bilgisayar programı aşamasında, kodun hata ayıklamasının yanı sıra tüm sayısal yöntemlerin gerçek kodlaması gerçekleştirilmektedir. Model aşamasında, model doğrulama ile ilgili tüm faaliyetler, yani deneysel verilerle karşılaştırmalar ve tahmin edilen sonuçların makul olup olmadığı kontrol edilmektedir. Modelleme sonucu aşamasında, sonuçların yorumlanması yapılır ve modelleme ve simülasyon çabasının orijinal amacına ulaşılmaya çalışılmaktadır. Tüm sürecin geri bildirimi ve yinelemeli doğası, modelleme ve simülasyon çabasını çevreleyen kesikli döngü ile gösterilmiştir (Şekil 3) (Jacoby ve ark., 1980).

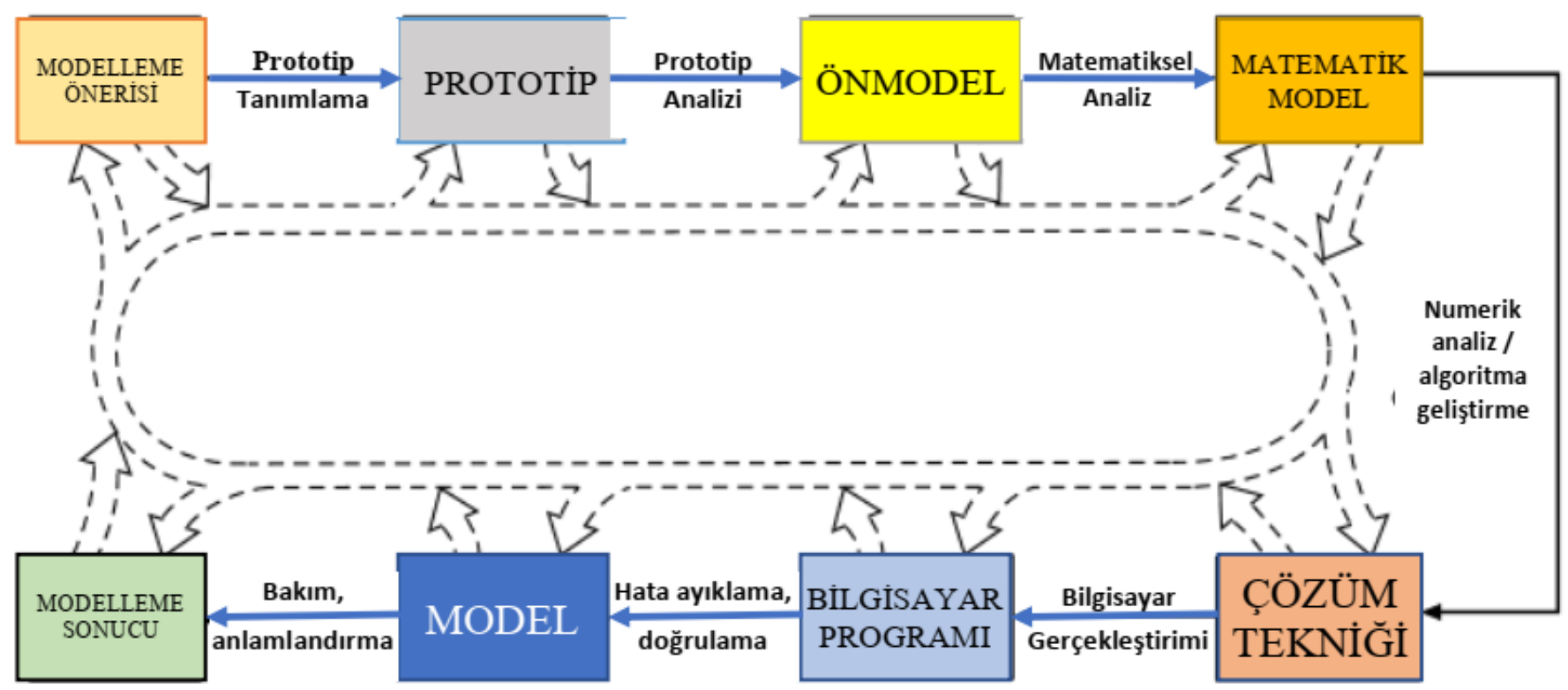

**Şekil 3:** Modelleme ve simülasyon aşamalarını gösteren tasarım (Jacoby ve Kowalik, 1980)

Sargent, 1980'lerdeki çalışmalarında (Sargent, 1980; Sargent, 1985), Şekil 2'de gösterilen modelleme ve simulasyon kavramlarını genelleştirmeye yönelik iyileştirmeler yapmıştır. Bu bağlamda en önemli katkısı, modellerin ve simulasyonların doğrulanması (*verificiation*) ve geçerlenmesi (*validation*) aşamaları için genel prosedürler geliştirmesi olmuştur.

Modelleme ve simulasyon aşamalarını daha ileri noktaya taşıyan Nance (1981) ve Balcı (1990), simulasyonu yaşam döngüsü kavramını içerecek şekilde tanımlamışlardır. Nance ve Balcı, önceki açıklamalara, 'Sistem ve Hedeflerin Tanımlanması', 'İletişimsel Modeller' ve 'Simulasyon Sonuçları' şeklinde majör aşamalar eklemişlerdir. 'Hedef Tanımlama' ve 'Simulasyon Sonuçları' aşamaları daha önce de (Jacoby ve ark., 1980) tanımlanmış olsa da uygulamada pek yer bulmamıştır. İletişimsel Modeller, Nance (1981) ve Balcı (1990) tarafından 'diğer insanlarla iletişim halinde olan, birden fazla insan tarafından sistem ve çalışma amacına karşı yargılanabilen veya mukayese edilebilen modeller' olarak tanımlanmıştır (Oberkampf ve ark., 2002).

## 2. SOSYAL BİLİMLERDE MODELLEME VE SİMÜLASYON

Sosyal bilimlerde bilgisayar simülasyonunun doğuşu zor olmuştur (Troitzsch, 1997). Daha önceki münferit uygulamalar olmasına rağmen, sosyal bilimlerde bilgisayar simülasyonu alanındaki ilk gelişmeler, 1960'ların başlarında bilgisayarların üniversite araştırmalarında ilk kez kullanımına denk gelmiştir (Şekil 1.4). Bu çalışmalar, esas olarak ayrık olay simülasyonlarından veya sistem dinamiklerine dayalı simülasyonlardan oluşmaktaydı. Bu alandaki çalışmaları şekillendiren ilk yaklaşımlar, tipik iş hacmini tahmin etmek için birimlerin kuyruklar ve süreçler üzerinden geçişini modellemekteydir. Müşterilerin kuyrukta bekleme süresinin veya bir şehrin polis arabalarının bir acil duruma ulaşması için geçen sürenin tahmin edilmesi gibi uygulamalar örnek verilebilir (Kolesar ve Walker 1975).

Sistem dinamiği yaklaşımında, zaman içindeki değişkenlerin yörüngelerini çizmek için büyük fark denklem sistemlerini kullanılmaktadır. Club of Rome'un dünya ekonomisinin geleceğine ilişkin çalışmaları bu yaklaşıma örnek olarak verilebilir (Meadows ve ark., 1974). Küresel çevre felaketini öngören Club of Rome simülasyonları büyük bir etki yaratmıştır fakat aynı zamanda sonuçların büyük ölçüde modelin parametreleri hakkında yapılan belirli nicel varsayımlara bağlı olduğu netleştiği için simülasyona kötü bir itibar kazandırmıştır. Bu nicel varsayımların çoğu yetersiz kanıtlarla desteklenmekteydi (Gilbert ve Troitzsch, 2005).

Birkaç yıl boyunca etkisini sürdürebilen "Simulmatics" (Sola Pool ve Abelson 1962) olarak tanınan bir yaklaşım öne sürülmüştür. Simulmatics projesi aslen John F. Kennedy'nin başkanlık kampanyasına danışmanlık yapmak için tasarlanmıştır. Bu proje ile, seçmenlerin Kennedy ve kampanya ekibi tarafından alınan önlemlere tepkileri tahmin edilmeye çalışılmış ve aynı zamanda 1960'ların başında Amerika Birleşik Devletleri'nde sık görülen içme suyunun florlanmasıyla ilgili referandum kampanyalarında seçmenlerin davranışlarını anlamak için kullanılmıştır (Abelson ve Bernstein, 1963).

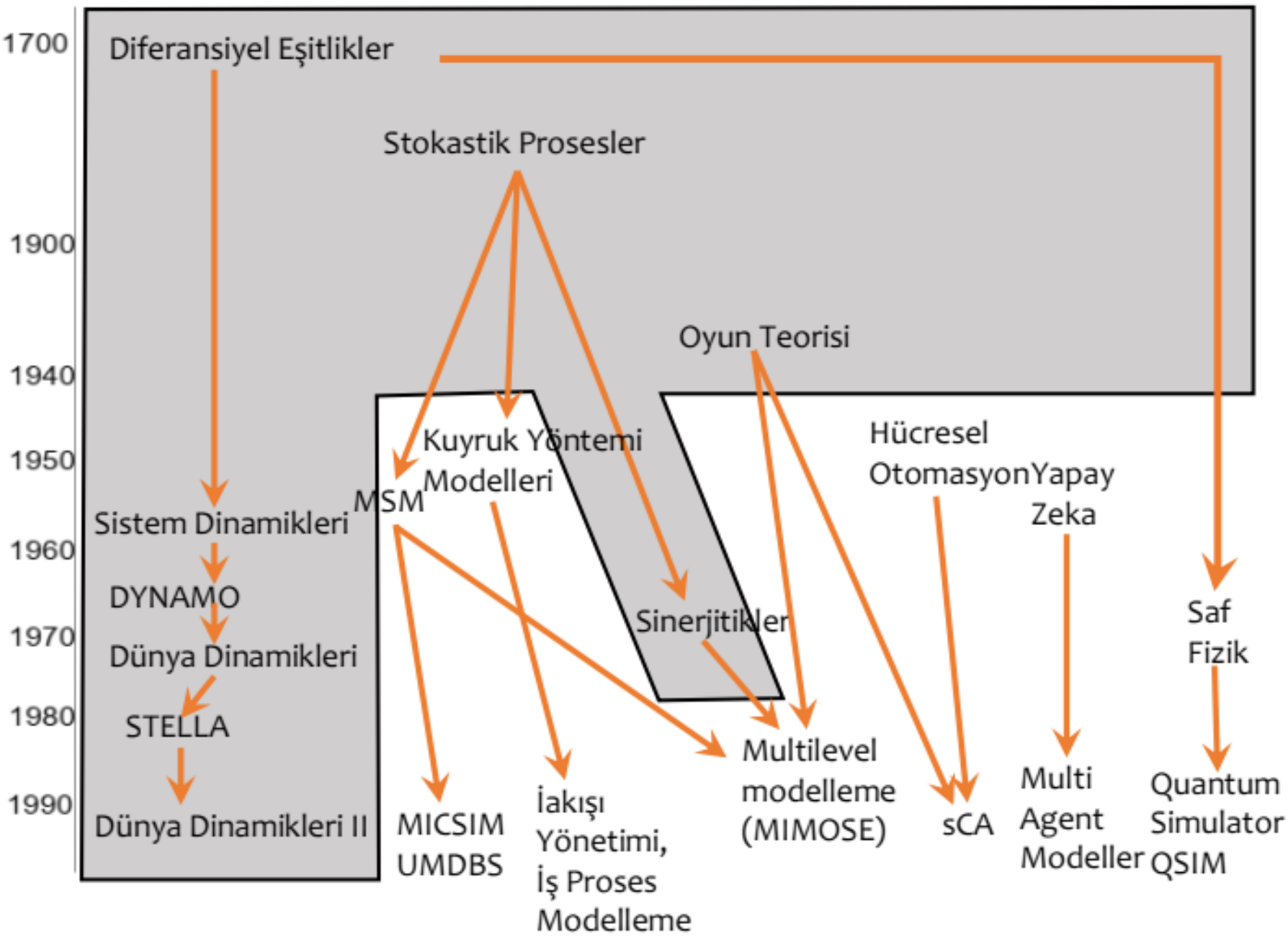

*Lejant: Gri alanlar: eşitlik temelli modeller; beyaz alanlar: nesne, olay ya da ajan temelli modeller. 'sCA' (cellular automata) hücresel otomasyon teriminin kısaltmasıdır.*

**Şekil 4:** Sosyal bilimlerde simülasyona çağdaş yaklaşımların gelişimi (Gilbert ve Troitzsch, 2005)

## 3. RİSK YÖNETİMİNDE MODELLEME VE SİMÜLASYON

Risk yönetim birimlerinde, özellikle nükleer reaktör güvenliği ve radyonüklidlerin çevresel etkisi konularında yürütülen çalışmalar, modelleme ve simülasyon fazlarını doğrudan ele almamıştır. Araştırmacılar, bunun yerine, risk değerlendirme tahminlerinde belirsizliğe katkıda bulunabilecek olası kaynaklara odaklanmışlardır. Reaktör güvenlik analizleri kapsamında risk değerlendirmesine katkı sunacak, olası arıza ve olay ağacı senaryolarını oluşturmak için kapsamlı yöntemler geliştirilmiştir (Morgan ve ark., 1990; Frank, 1999; Ang ve De Leon, 2005; Hauptmanns, 2012).

Düşük ve yüksek seviyeli nükleer atıkların bertarafı için jeolojik depoların risk analizleri yapılırken, senaryo analizleri kullanılmış ve risk analizinin diğer aşamalarında meydana gelen belirsizlik ve yanlışlık kaynakları belirlenmiştir. Spesifik olarak, bu analizler kavramsal modellemede, matematiksel modellemede, bilgisayar kodu uygulamasında ve deneysel olarak ölçülen veya türetilmiş model girdi verilerinde meydana gelen farklı türde kaynakları tanımlamıştır (Davis ve ark., 2010; Davis ve ark, 1990).

## 4. BULUT TABANLI BİLİŞİM SİSTEMLERİNDE MODELLEME VE SİMULASYON

Bulut bilişim, bilgi işlem kaynaklarını bir ağ üzerinden bir hizmet olarak sağlar. Gelişen bu teknolojinin gerçek dünyada hızlı bir şekilde uygulanması ile bulut bilişimin karşılaştığı performans ve güvenlik sorunlarının nasıl değerlendirileceği giderek daha fazla önem kazanmaktadır. Çok yakın zamanda, bulut bilişimin ve bu platformda yürütülen uygulamaların nasıl performans göstereceğini, bulut bilişim hizmetlerinin güvenli ve gizliliği koruyup korumadığını ve hangi bulut bilişim hizmeti kullanıcılarının seçebileceğini incelemek ve analiz etmek giderek daha önemli hale gelmektedir (Zhao ve ark., 2012).

Halihazırda modelleme ve simülasyon teknolojisi, bulut bilişim araştırma topluluğunda bu sorunlarla başa çıkmak için kullanışlı ve güçlü bir araç haline gelmiştir. Bulut bilişimin modellemesi ve simülasyonu ile ilgili mevcut sonuçların gözden geçirildiği bir çalışmada iki tür bulut bilişim simülatörü olduğu belirtilmiştir. Bu simülatörler, sadece yazılıma dayalı simülatörler ve hem yazılım hem de donanıma dayalı simülatörler olarak sınıflandırılmaktadır (Zhao ve ark., 2012). Gerçek bir ortamda gerçekleştirilecek denemelerin pahalı, zaman maliyetli ve tekrarlanamaz olmasından dolayı, gerçek bulut ortamlarındaki performans ve güvenlik sorunlarını analiz etmek genellikle zordur (Calheiros ve ark., 2013). Bu nedenle, modelleme ve simülasyon teknolojisi, bulut endüstrisinde ve akademide giderek daha popüler hale gelmektedir. Modelleme ve simülasyon teknolojisinin, fiziksel dünyadaki karmaşık sorunları analiz etmek için sıklıkla kullanılan bir yöntem olduğu iyi bilinmektedir, simülasyon tabanlı deneylerin çoğu, bulut bilişim ve uygulama ortamlarının basitleştirilmiş modellemesini dikkate alır (Wang ve ark., 2011).

Literatürde bazı geleneksel dağıtık sistem simülatörleri (Buyya & Murshed, 2002; Dumitrescu & Foster, 2005; Legrand ve ark., 2003) bulut bilişim bileşenleri tarafından doğrudan kullanılabilecek ortamı sağlayamamaktadır. Bu nedenle, son yıllarda piyasaya sürülen özellikle bulut bilişime özgü simülatörler ortaya çıkmıştır. Bunlar, CloudSim (Calheiros ve ark., 2011), The Open Cloud Testbed (Grossman ve ark., 2009), GreenCloud (Kliazovich ve ark., 2012), iCanCloud (Nunez ve ark., 2011), SPECI (Sriram, 2009), Cloud Analysis (Buyya, 2009) şeklinde listelenebilir.

## 5. AJAN TABANLI MODELLEME VE SİMÜLASYON

Pek çok durum ve alt sistemin nasıl çalışacağı, bir dizi özerk yapıların davranışlarına sıkı sıkıya bağlı olabilir. Ajan tabanlı(*agent-based*) modeller özellikle bu durumların üstesinden gelmek için uygundur. Bu modeller, merkezi olmayan karar verme, lokal-global etkileşimler, kendi kendine organizasyon, simüle edilmiş sistemde heterojenliğin ortaya çıkması ve etkileri gibi konuların incelenmesini ve analizini desteklemektedir.

Weyns ve ark. (2005), çalışmalarında lojistik optimizasyondan çok farklı bağlamlarda karmaşık sistemleri simüle etmek amacıyla, (belirli bir dereceye kadar) otonom ajanlara dayalı modelleri başarıyla tasarlamışlardır. Çalışmalarında, çok etmenli (*multi-agent*) modelleme ve simulsayon proseslerindeki temel kavramları ve mekanizmaları araştırmak, denemek ve değerlendirmek amacıyla Packet-World adlı soyut bir uygulamayı kullanılmıştır. Aktif algılama, yerleşik ajanların karar vermesi, eşzamanlı eylemlerin senkronizasyonu ve dolaylı koordinasyon gibi olayların örnekleri oluşturulmuştur. Packet-World, insansız araçlar aracılığıyla bir depo taşıma sisteminin merkezi olmayan kontrolü gibi gerçek dünya uygulamalarının benzetimi şeklinde düşünülmüştür (Weyns ve ark., 2005).

Bandini ve ark. (2005), çok etmenli modelleme ve simulayonu, bağışıklık sistemini (*Immune System* - IS) ilgilendiren biyolojik bir çalışmada uygulamışlardır. Bağışıklık sisteminin, heterojen özerk varlıkların işbirliği yoluyla bilinmeyen tehditlere uyum sağlama mekanizmalarıyla donatılmış, dağıtılmış bir sistem olma fikri çalışmalarının motivasyonunu oluşturmuştur. Çalışmalarında, çoklu ajan yaklaşımının ve daha spesifik olarak Yerleşik Hücresel Ajanlar (*Situtated Cellular Agent* - SCA) modelinin, bağışıklık sisteminin belirli unsurlarını ve mekanizmalarını temsil etmek için uygun şekilde nasıl tasarlanabileceğini açıklamışlardır. Bağışıklık sistemini oluşturan parçaların ve iç mekanizmalarının kısa bir açıklamasından sonra, SCA modeli tanıtılmış ve simulasyon sonuçları paylaşılmıştır (Bandini ve ark., 2005).

Bu uygulamalardan farklı olarak, ajan temelli modelleme ve simülasyon yaklaşımı, trafik uygulamalarında (Bazzan ve ark., 1999; Wahle ve Schreckenberg, 2001; Balmer ve Nagel, 2006), yaya simulasyonunda (Batty, 2001), şehir ve bölge planlama uygulamasında (Arentze ve Timmermans, 2003), sosyal bilimlerde (Epstein, 1999) ve eknomoi alanındaki çalışmalarda (Lane, 1993; Windrum ve ark. 2007) başarıyla uygulanmıştır.

## KAYNAKÇA

Abelson, R. P., & Bernstein, A. (1963). A computer simulation model of community referendum controversies. *Public Opinion Quarterly*, *27*(1), 93-122.

Ang, A. S., & De Leon, D. (2005). Modeling and analysis of uncertainties for risk-informed decisions in infrastructures engineering. *Structure and infrastructure engineering*, *1*(1), 19-31.

Arentze, T., & Timmermans, H. (2003). A multiagent model of negotiation processes between multiple actors in urban developments: a framework for and results of numerical experiments. *Environment and planning B: planning and design*, *30*(3), 391-410.

Balci, O. (1990). *Guidelines for successful simulation studies*. Institute of Electrical and Electronics Engineers (IEEE).

Balmer, M., & Nagel, K. (2006). Shape morphing of intersection layouts using curb side oriented driver simulation. In *Innovations in design & decision support systems in architecture and urban planning* (pp. 167-183). Springer, Dordrecht.

Bandini, S., Celada, F., Manzoni, S., Puzone, R., & Vizzari, G. (2005). Modelling the immune system with situated agents. In *Neural Nets* (pp. 231-243). Springer, Berlin, Heidelberg.

Bandini, S., Manzoni, S., & Vizzari, G. (2009). Agent based modeling and simulation: an informatics perspective. *Journal of Artificial Societies and Social Simulation*, *12*(4), 4.

Banks, J. (Ed.). (1998). *Handbook of simulation: principles, methodology, advances, applications, and practice*. John Wiley & Sons.

Batty, M. (2001). Editorial: Agent-based pedestrian modeling. *Environment and Planning B: Planning and Design*, *28*(3), 321-326.

Bazzan, A. L. C., Wahle, J., & Klügl, F. (1999, September). Agents in traffic modelling—from reactive to social behaviour. In *Annual Conference on Artificial Intelligence* (pp. 303-306). Springer, Berlin, Heidelberg.

Bossel, H. (2018). *Modeling and simulation*. AK Peters/CRC Press.

Buyya, R. (2009). CloudAnalyst: A CloudSim-based tool for modelling and analysis of large scale cloud computing environments. *Distrib. Comput. Proj. Csse Dept., Univ. Melb*, 433-659.

Buyya, R., & Murshed, M. (2002). Gridsim: A toolkit for the modeling and simulation of distributed resource management and scheduling for grid computing. *Concurrency and computation: practice and experience*, *14*(13-15), 1175-1220.

Calheiros, R. N., Netto, M. A., De Rose, C. A., & Buyya, R. (2013). EMUSIM: an integrated emulation and simulation environment for modeling, evaluation, and validation of performance of cloud computing applications. *Software: Practice and Experience*, *43*(5), 595-612.

Calheiros, R. N., Ranjan, R., Beloglazov, A., De Rose, C. A., & Buyya, R. (2011). CloudSim: a toolkit for modeling and simulation of cloud computing environments and evaluation of resource provisioning algorithms. *Software: Practice and experience*, *41*(1), 23-50.

Davis, P. A., Bonano, E. J., Price, L. L., & Wahi, K. K. (1990). *Uncertainties Associated with performance assessment of high-level radioactive waste repositories* (No. NUREG/CR--5211). Nuclear Regulatory Commission.

Davis, P. A., Price, L. L., Wahi, K. K., Goodrich, M. T., Gallegos, D. P., Bonano, E. J., & Guzowski, R. V. (1990). *Components of an overall performance assessment methodology* (No. NUREG/CR-5256; SAND-88-3020; TI-90-008483). US Nuclear Regulatory Commission (NRC), Washington, DC (United States). Division of High-Level Waste Management; GRAM, Inc., Albuquerque, NM (United States); Science Applications International Corp., Albuquerque, NM (United States); Sandia National Lab.(SNL-NM), Albuquerque, NM (United States).

de Sola Pool, I., & Abelson, R. (1961). The simulmatics project. *Public Opinion Quarterly*, *25*(2), 167-183.

Epstein, J. M. (1999). Agent-based computational models and generative social science. *Complexity*, *4*(5), 41-60.

Frank, M. V. (1999). Treatment of uncertainties in space nuclear risk assessment with examples from Cassini mission applications. *Reliability Engineering & System Safety*, *66*(3), 203-221.

Gilbert, N., & Troitzsch, K. (2005). *Simulation for the social scientist*. McGraw-Hill Education (UK).

Grossman, R., Gu, Y., Sabala, M., Bennet, C., Seidman, J., & Mambratti, J. (2009). The open cloud testbed: A wide area testbed for cloud computing utilizing high performance network services. *arXiv preprint arXiv:0907.4810*.

Hauptmanns, U., & Werner, W. (2012). *Engineering risks: Evaluation and valuation*. Springer Science & Business Media.

Ho, Y. C. (Ed.). (1992). *Discrete event dynamic systems: analyzing complexity and performance in the modern world*. IEEE.

Jacoby, S. L., Kowalik, J. S., & Burner, H. B. (1980). *Mathematical modeling with computers*. Prentice Hall.

Kliazovich, D., Bouvry, P., & Khan, S. U. (2012). GreenCloud: a packet-level simulator of energy-aware cloud computing data centers. *The Journal of Supercomputing*, *62*(3), 1263-1283.

Kolesar, P., & Walker, W. E. (1975). A simulation model of police patrol operations: program description.

Lane, D. A. (1993). Artificial worlds and economics, part I. *Journal of evolutionary economics*, *3*(2), 89-107.

Law, A. M., Kelton, W. D., & Kelton, W. D. (2000). *Simulation modeling and analysis* (Vol. 3). New York: McGraw-Hill.

Legrand, A., Marchal, L., & Casanova, H. (2003, May). Scheduling distributed applications: the simgrid simulation framework. In *CCGrid 2003. 3rd IEEE/ACM International Symposium on Cluster Computing and the Grid, 2003. Proceedings.* (pp. 138-145). IEEE.

Meadows, D. L., Behrens, W. W., Meadows, D. H., Naill, R. F., Randers, J., & Zahn, E. (1974). *Dynamics of growth in a finite world* (p. 637). Cambridge, MA: Wright-Allen Press.

Morgan, M. G., Henrion, M., & Small, M. (1990). *Uncertainty: a guide to dealing with uncertainty in quantitative risk and policy analysis*. Cambridge university press.

Nance, R. E. (1981). Model representation in discrete event simulation: the conical methodology.

Neelamkavil, F. (1987). *Computer simulation and modelling*. John Wiley & Sons, Inc..

Nunez, A., Vazquez-Poletti, J. L., Caminero, A. C., Carretero, J., & Llorente, I. M. (2011, June). Design of a new cloud computing simulation platform. In *International Conference on Computational Science and Its Applications* (pp. 582-593). Springer, Berlin, Heidelberg.

Oberkampf, W. L., DeLand, S. M., Rutherford, B. M., Diegert, K. V., & Alvin, K. F. (2002). Error and uncertainty in modeling and simulation. *Reliability Engineering & System Safety*, *75*(3), 333-357.

Sargent, R. G. (1984). Simulation model validation. In *Simulation and model-based methodologies: an integrative view* (pp. 537-555). Springer, Berlin, Heidelberg.

Sargent, R. G. (1985, December). An expository on verification and validation of simulation models. In *Proceedings of the 17th conference on Winter simulation* (pp. 15-22).

Schlesinger, S. (1979). Terminology for model credibility. *Simulation*, *32*(3), 103-104.

Sriram, I. (2009, December). SPECI, a simulation tool exploring cloud-scale data centres. In *IEEE International Conference on Cloud Computing* (pp. 381-392). Springer, Berlin, Heidelberg.

Troitzsch, K. G. (1997). Social science simulation—origins, prospects, purposes. In *Simulating social phenomena* (pp. 41-54). Springer, Berlin, Heidelberg.

Vesely, W. E., Goldberg, F. F., Roberts, N. H., & Haasl, D. F. (1981). *Fault tree handbook*. Nuclear Regulatory Commission Washington DC.

Wang, Q., Ren, L., & Zhang, L. (2011, June). Design and implementation of virtualization-based middleware for cloud simulation platform. In *4th International conference on Computer Science and information Technology* (pp. 10-12).

Wahle, J., & Schreckenberg, M. (2001, January). A multi-agent system for on-line simulations based on real-world traffic data. In *Proceedings of the 34th annual Hawaii international conference on system sciences* (pp. 9-pp). IEEE.

Weyns, D., Helleboogh, A., & Holvoet, T. (2005). The packet-world: A test bed for investigating situated multi-agent systems. In *Software Agent-Based Applications, Platforms and Development Kits* (pp. 383-408). Birkhäuser Basel.

Windrum, P., Fagiolo, G., & Moneta, A. (2007). Empirical validation of agent-based models: Alternatives and prospects. *Journal of Artificial Societies and Social Simulation*, *10*(2), 8.

Zeigler, B. P., Muzy, A., & Kofman, E. (2018). *Theory of modeling and simulation: discrete event & iterative system computational foundations*. Academic press.

Zhao, W., Peng, Y., Xie, F., & Dai, Z. (2012, November). Modeling and simulation of cloud computing: A review. In *2012 IEEE Asia Pacific cloud computing congress (APCloudCC)* (pp. 20-24). IEEE.